\documentclass[10pt]{IEEEtran}
\IEEEoverridecommandlockouts

\usepackage{cite}
\usepackage{amsmath,amssymb,amsfonts}
\usepackage{algorithmic}
\usepackage{algorithm}
\usepackage{lipsum}
\usepackage{graphicx}
\usepackage{epsfig}
\usepackage{amsmath}
\usepackage{bbm}
\usepackage{booktabs}
\usepackage{multirow}
\usepackage[normalem]{ulem}
\ifCLASSOPTIONcompsoc
    \usepackage[caption=false, font=normalsize, labelfont=sf, textfont=sf]{subfig}
\else
\usepackage[caption=false, font=footnotesize]{subfig}
\fi
\usepackage{textcomp}
\usepackage{xcolor}
\usepackage{verbatim}

\def\BibTeX{{\rm B\kern-.05em{\sc i\kern-.025em b}\kern-.08em
    T\kern-.1667em\lower.7ex\hbox{E}\kern-.125emX}}
\usepackage[nopostdot,nogroupskip,style=super,nonumberlist,acronym]{glossaries}
\newacronym{aoi}{AoI}{Age of Information}
\newacronym{aodv}{AODV}{Ad hoc On-demand Distance Vector}
\newacronym{cnn}{CNN}{convolutional neural networks}
\newacronym{deepl}{DL}{Deep Learning}
\newacronym{dod}{DoD}{depth of discharge}
\newacronym{dqn}{DQN}{deep Q-learning}
\newacronym{gsl}{GSL}{ground-to-satellite link}
\newacronym{isl}{ISL}{inter-satellite link}
\newacronym{leo}{LEO}{low Earth orbit} 
\newacronym{ml}{ML}{Machine Learning}
\newacronym{mdp}{MDP}{Markov decision process}
\newacronym{ngeo}{NGSO}{Non-geostationary orbit}
\newacronym{ngso}{NGSO}{Non-geostationary orbit}
\newacronym{olsr}{OLSR}{optimized link state routing protocol}
\newacronym{ospf}{OSPF}{Open Shortest Path First}
\newacronym{pan}{PAN}{Path-Aware Networking}
\newacronym{qos}{QoS}{Quality of Service}
\newacronym{rl}{RL}{Reinforcement Learning}
\newacronym{drl}{DRL}{Deep \gls{rl}}
\newacronym{dnn}{DNN}{Deep Neural Network}
\newacronym{dql}{DQL}{Deep Q-learning}
\newacronym{ql}{QL}{Q-learning}
\newacronym{e2e}{E2E}{end-to-end}
\newacronym{bgp}{BGP}{Border Gateway Protocol}
\newacronym{ibgp}{iBGP}{interior Border Gateway Protocol}
\newacronym{ebgp}{eBGP}{exterior Border Gateway Protocol}
\newacronym{as}{AS}{Autonomous System}
\newacronym{relu}{ReLu}{Rectified Linear Unit}
\newacronym{cdf}{CDF}{Cumulative Distribution Function}
\newacronym{ue}{UE}{User Equipment}
\newacronym{gps}{GPS}{Global Positioning System}
\newacronym{pomdp}{POMDP}{Partially Observable \gls{mdp}}
\newacronym{snr}{SNR}{signal-to-noise ratio}
\newacronym{awgn}{AWGN}{additive-white Gaussian noise}
\newacronym{bm}{BM}{benchmark}
\newacronym{fifo}{FIFO}{first-in first-out}
\newacronym{ip}{IP}{Internet Protocol}
\newacronym{gsd}{GSD}{Ground Sample Distance}
\newacronym{eo}{EO}{Earth Observation}
\newacronym{iot}{IoT}{Internet of Things}
\newacronym{semcom}{SemCom}{Semantic Communications}
\newacronym{ssim}{SSIM}{structural similarity index measure}
\newacronym{sdg}{SDG}{Sustainable Development Goal}
\newacronym{ai}{AI}{Artificial Intelligence}
\newacronym{mse}{MSE}{Mean Square Error}
\newacronym{fid}{FID}{Fréchet Inception Distance}
\newacronym{jscc}{JSCC}{Joint Source and Channel Coding}
\newacronym{kg}{KG}{Knowledge Graph}
\newacronym{gtfp}{GTFP}{ground track frame period}
\newacronym{gs}{GS}{ground station}
\newacronym{fov}{FOV}{Field of View}
\newacronym{fso}{FSO}{free-space optical}
\newacronym{ntn}{NTN}{Non-Terrestrial Networks}
\newacronym{lsatc}{LSatC}{Low Earth Orbit Satellite Constellations}
\newacronym{fm}{FM}{Foundation Model}
\newacronym{genai}{GenAI}{Generative AI}
\newacronym{ul}{UL}{uplink}
\newacronym{dl}{DL}{downlink}
\newacronym{nir}{NIR}{near infrared}
\newacronym{uv}{UV}{ultraviolet}
\newacronym{map}{mAP}{mean Average Precision}
\newacronym{fps}{FPS}{frames per second}
\newacronym{kb}{KB}{Knowledge Base}
\newacronym{soa}{SoA}{State-of-the-Art}
\newacronym{cpu}{CPU}{Central Processing Unit}
\newacronym{rf}{RF}{Radio Frequency}
\newacronym{gpu}{GPU}{Graphics Processing Unit}
\newacronym{vnir}{VNIR}{Visible and Near Infra-Red}
\newacronym{swir}{SWIR}{Short Wave Infra-Red}
\newacronym{mwir}{MWIR}{Mid Wave Infra-Red}
\newacronym{lwir}{LWIR}{Long Wave Infra-Red}

\usepackage{comment}

\title{Semantic and goal-oriented edge computing for satellite Earth Observation}
\author{Beatriz Soret, Israel Leyva-Mayorga, Antonio M. Mercado-Martínez, Marco Moretti, Antonio Jurado-Navas, Marc Martinez-Gost, Celia Sánchez de Miguel, Ainoa Salas-Prendes, and Petar Popovski

\vspace{-1cm}

\thanks{B. Soret (bsoret@ic.uma.es), A. M. Mercado-Martínez, C. Sánchez de Miguel, and A. Jurado-Navas are with the Telecommunications Research Institute, University of Malaga, Malaga, Spain. I. Leyva-Mayorga, A. Salas-Prendes and P. Popovski are with the Connectivity Section, Aalborg University, Aalborg, Denmark. Marco Moretti is with the University of Pisa, Pisa, Italy. Marc Martinez-Gost is with the Centre Tecnològic de Telecomunicacions de Catalunya (CTTC) and Polytechnic University of Catalonia, Barcelona, Spain. This work is partially funded by ESA SatNEx V (prime contract no. 4000130962/20/NL/NL/FE), and by the Spanish Ministerio de Ciencia, Innovación y Universidades (''TATOOINE'', grant no. PID2022-136269OB-I00). The view expressed herein can in no way be taken to reflect the official opinion of the European Space Agency. 
}
}

\begin{document}
\bstctlcite{IEEEexample:BSTcontrol}
\maketitle

\begin{abstract}

The integration of \gls{semcom} and edge computing in space networks enables the optimal allocation of the scarce energy, computing, and communication resources for data-intensive applications. We use \gls{eo} as a canonical functionality of satellites and review its main characteristics and challenges. We identify the potential of the space segment, represented by a \gls{leo} satellite constellation, to serve as an edge layer for distributed intelligence. Based on that, propose a system architecture that supports semantic and goal-oriented applications for image reconstruction and object detection and localization. 
The simulation results show the intricate trade-offs among energy, time, and task-performance using a real dataset and \gls{soa} processing and communication parameters. 


\vspace{-0.4cm}

\end{abstract}

\IEEEpeerreviewmaketitle

 \glsresetall
\section{Introduction}
Resource allocation is a major challenge in satellite networks given the constraints in the energy, computing, and communication subsystems. A set of satellite-based applications that fall in the data-intensive category is \gls{eo}, which provide crucial information for climate and environmental monitoring, maritime surveillance, or disaster management, among others.
The vast amount of collected data, as well as the limited contact time and capacity of the feeder link, connecting the ground segment and the space segment, make it infeasible to have a brute-force data transmission for real-time \gls{eo}. This situation is about to change due to the synergy between new technology enablers and the redefinition of application objectives and requirements. 

New technology enablers include three components. \emph{(i)} \gls{ai} and Deep Learning, leading to remarkable progress in data processing, including image and video, capable of transforming raw data into actionable intelligence. \emph{(ii)} The standardization of \gls{ntn} in 5G/6G by 3GPP, with \gls{leo} satellite constellations as the flagship infrastructure providing new computation and communication capabilities. In addition, the small satellite business is growing and drastically reducing the cost of the missions through specialized yet affordable spacecrafts and innovative service models. 
\emph{(iii)} Edge computing to process and analyze information closer to the source, resulting in reduced latency and bandwidth, while enabling faster, more efficient decision-making. With \gls{leo} satellite networks and their computing resources, the edge will 
process the raw \gls{eo} data before being sent to the ground, de-congesting the communication network~\cite{Ley23TCOM}. 

The following three applications, with increasing level of complexity, illustrate the redefinition of application objectives and requirements. \emph{(i)} \emph{Image reconstruction}: A ground user is interested in getting images from a given area of the Earth's surface. The  images should be compressed in space in a way that satisfies certain distortion criteria. \emph{(ii)} \emph{Real-time object detection and localization}: A ground user is interested in a real-time object mapping. Instead of sending the raw data, the satellites run an object detection and localization algorithm on the acquired images and encode the result. \emph{(iii)} \emph{Real-time object tracking with closed-loop control of the satellite:} A ground monitor tracks moving objects in real time, where a remote closed-loop control decides the orbit/altitude parameters for data acquisition based on the status of the network, weather, or tracked physical entities. 

These applications illustrate the transformation of requirements from sending raw data towards sending small portions of it that are compressed according to its application significance or goal. This requires a careful interplay between the technology enablers in terms of computation, \gls{ai}, and communication, to attain the application objectives within the given time, energy, and accuracy targets. As such, the described applications fall under the umbrella of \emph{\gls{semcom}}~\cite{deniz2023semantic}\cite{Luo2022semanticmagazine}, factoring the meaning and significance of the information into the communication process. 

\begin{figure}[t]
\centering
\includegraphics[width=\columnwidth] {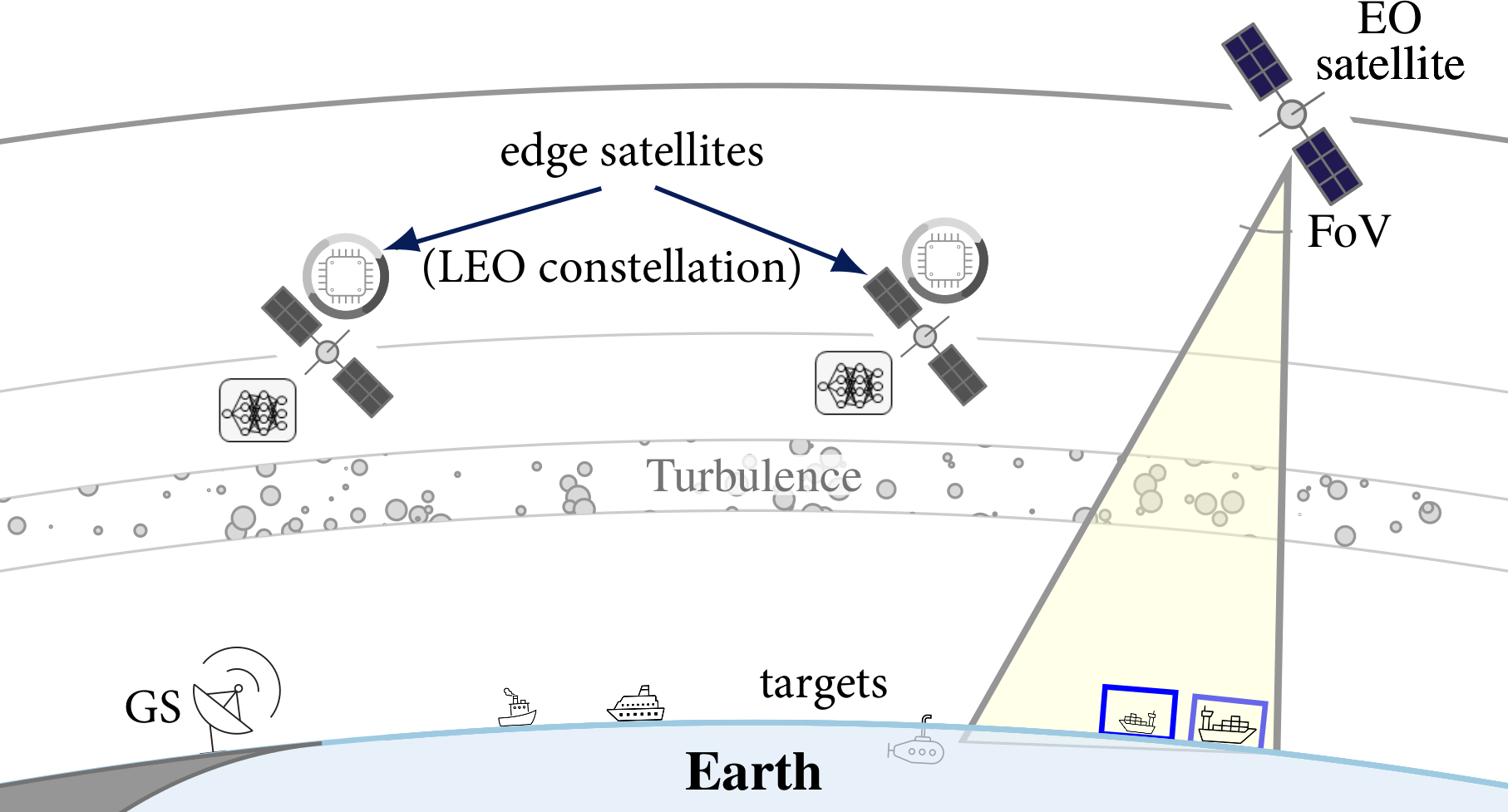}
\caption{Sketch of \gls{eo} vessel detection and localization. \vspace{-0.5cm}
} \label{fig:scenario}
\end{figure}

This paper advocates the use of the \gls{semcom} framework in the context of satellite computing and communication. We elaborate upon the fundamental concepts and their instantiation via technology enablers. Edge computing is instrumental for \gls{semcom} by supporting the semantic encoding/decoding process, providing the required computing and storage resources, while adhering to the  bandwidth constraints. Figure~\ref{fig:scenario} shows how a generic \gls{eo} application is supported by \gls{semcom}.
\gls{eo} satellites(s) capture(s) pictures of the area of interest, which are subject to atmospheric turbulence. These \gls{eo} satellite(s) may be part of, or separate from, the \gls{leo} satellite constellation, which serves as the edge layer that enhances the computing capabilities of ground-based systems. The processed information is transmitted to the ground for final aggregation, interpretation, and presentation of the results. 

\section{Fundamentals of Earth Observation}

 \gls{eo} satellites generate large amounts of  data about the Earth's surface, water bodies, and atmosphere. For example, the ESA Sentinel missions acquire approximately 12 TB of images daily, while NASA missions collectively provide \mbox{20-30 TB} per day. These volumes are rapidly increasing as new missions are planned and launched. \gls{eo} images are used for different purposes requiring computer vision processing and inference, e.g., object localization and detection, semantic labeling of image regions, change detection, or time series analysis. Table ~\ref{tab:soa} provides a representative list of \gls{eo} missions and \gls{soa} processing algorithms. 

\subsection{Data acquisition}  
The \gls{eo} satellite payload includes various remote sensing instruments, such as cameras, radar, spectrometers, or thermal sensors. 
The optical cameras are often multi-spectral, ranging from the \gls{nir} to the \gls{uv} wavelengths. A frame is typically obtained by superimposing frames captured at different wavelengths. A color frame is produced by combining images taken at red, green, and blue (RGB) wavelengths, often termed \emph{panchromatic} image. 

The area covered and the quality of the images are primarily affected by \emph{(i)} the altitude of the orbit; \emph{(ii)} the \gls{fov}, which is the angular width that determines the extent of the observable area captured by the camera sensor; \emph{(iii)} the \gls{gsd}, which is the actual distance on the ground between the centers of two adjacent pixels in the image, determining the spatial resolution of the imagery~\cite{Ley23TCOM}.  Typically, lower altitudes correspond to smaller values of \gls{gsd} and \gls{fov}. Since high image definition requires small \gls{gsd} values, \gls{leo} satellites are the prevalent choice for \gls{eo}. As the captured frame extends further off-nadir (away from the point directly below the satellite), the \gls{gsd} increases, resulting in lower resolution. Therefore, the satellite's coverage area needs to be precisely defined to ensure that images are captured with the desired quality. For this, the attitude subsystem of an \gls{eo} satellite controls and adjusts the orientation of the spacecraft along the three axes (roll, pitch, and yaw),
with new generation \gls{eo} agile remote sensing satellites~\cite{Wang_2021} being capable of self-tuning of the orbit and covered area. 
Furthermore, \emph{pansharpening}, where a multispectral frame is combined with a panchromatic frame, can be used to reduce the \gls{gsd} and enhance image quality.


\begin{table*}[t]
\caption{Earth Observation (EO) missions and \gls{soa} image processing algorithms.}

\centering
\renewcommand{\arraystretch}{1.2}
\begin{tabular}{@{} p{2.5cm} p{2.8cm} p{5.5cm} p{5.7cm}@{}}
\toprule
\multicolumn{4}{c}{\textbf{EO missions}}\\
\midrule
\textbf{Name}& \textbf{Agency/Company}& \textbf{Constellation} & \textbf{Imaging capabilities}\\\midrule
Sentinel-2 & ESA & 2  satellites at 786 km. &13 spectral bands from \gls{vnir} to \gls{swir}.\\
GOES & NOAA and NASA & 3 geostationary satellites. & 16 spectral bands: \gls{vnir}, \gls{mwir}, and \gls{lwir}. \\
PlanetScope \cite{PlanetScope} & PlanetLabs & $>430$ 3U CubeSats Doves and SuperDoves at 475-525 km. & 8-band multispectral images. \\
WorldView Series \cite{worldview-13} 
  & Maxar & 3  satellites at 496 km (WorldView-1), 770 km (WorldView-2) and 617 km (WorldView-3)&  WorldView-1: panchromatic images. WorldView-2 and 3: 8-band multispectral  images.\\
  \midrule
  \multicolumn{4}{c}{\textbf{Image processing algorithms}}\\
  \midrule
  \textbf{Name}& \textbf{Application}& \textbf{Architecture and characteristics} & \textbf{Performance}\\\midrule
  JPEG2000 & Compression & Scalable lossy and lossless compression. & --\\
  Single Shot Detector (SSD) 
& Object detection and localization& Single CNN to predict multiple bounding boxes and their class probabilities from multiple feature maps. & \Gls{map} @0.5 $\geq 68\%$ on PASCAL VOC2007 dataset.\\
YOLO (v8) \cite{yolov8_ultralytics} 
&  Classification, segmentation, object detection, and tracking& Multiple CNN and fully connected layers to predict the bounding boxes of the objects & \gls{map} @0.5 = $76.8\%$.\\
GOTURN \cite{held2016learningtrack100fps} & Time series analysis& CaffeNet architecture to predict the bounding box coordinates of an object in the frame from a previous one. & Accuracy = 61\%, Robustness = 90\%.\\
EndNet \cite{Hong_EndNet_2022} & Multimodal data fusion& Encoder-decoder NN & Averaged accuracy = $93.88\%$.\\
\bottomrule
\end{tabular}

\label{tab:soa}

\end{table*}



\begin{figure*}[t]
\centering
\includegraphics{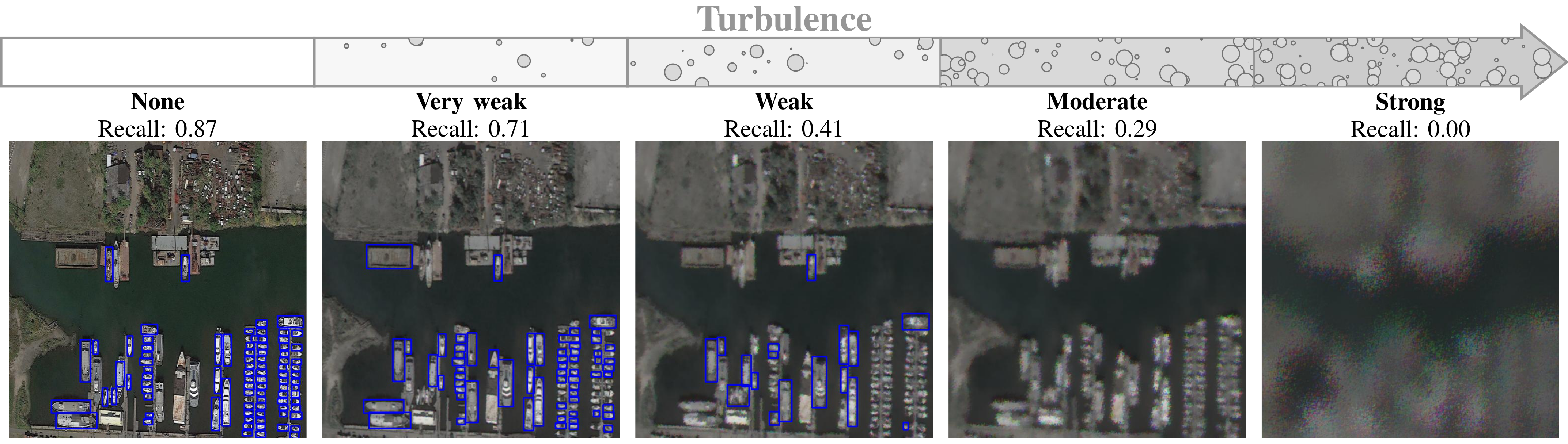}
\caption{Impact of the atmospheric turbulence in ship detection and localization using YOLOv8. The output of the algorithm are the blue bounding boxes that include coordinates and confidence about the detection and its performance is measured by the \emph{recall}: the ratio of correctly detected objects to all actual objects. }
\vspace{-0.4 cm}\label{fig:atmospheric}
\end{figure*}

\subsection{Meteorological effects}



Meteorological factors are essential in applying image acquisition and processing techniques. The atmosphere constitutes a massive thermodynamic system  
with slow, periodic oscillations caused by atmospheric tides, and small fluctuations in the atmospheric refractive index due to slight temperature variations. 
The thermal micro-variations are explained by the appearance of convective air currents caused, horizontally, by the progressive warming of the ground by the Sun and, vertically, by velocity gradients between different atmospheric layers. The latter modifies the uniform characteristics of medium's viscosity due to dynamic mixing and random subflows, called turbulent vortices or eddies \cite{andrews2005laser},
which induce characteristic image aberrations, compromising the achievable angular camera resolution.  

Atmospheric turbulence can be modelled through phase screens, which contain the phase fluctuation related to the refractive index fluctuation spectrum. 
Typically, a thin phase screen technique is employed based on the variation imposed by the Kolmogorov power spectrum density of the index of \hbox{refraction \cite{andrews2005laser}.} This phase screen is then superimposed on the wavefront of the light as it propagates through the simulated atmospheric layer, allowing realistic simulation of how  lightwave aberrations are captured by the camera. Figure \ref{fig:atmospheric} shows an example of the impact of the atmospheric turbulence in the performance of a given object detection and localization algorithm, YOLOv8, applied to a pre-trained \gls{eo} database~\cite{ijgi11080445}. As the turbulence becomes more severe, parameterized in larger refractive index fluctuations, the algorithm struggles to do the vessel detection.

\subsection{Data processing} 
The acquired images can be processed and compressed using various algorithms for compression, ranging from traditional methods like JPEG to advanced \gls{ai} algorithms designed for specific tasks. The selection of the specific algorithm depends on multiple factors, including the compression factor, algorithm complexity, and task accuracy. The compression factor indicates the reduction of the data size,
which is critical for efficient data storage and transmission. The algorithm complexity affects the time elapsed from data capture until the information is ready for transmission, as well as the energy consumed during processing. More complex algorithms require additional resources but may provide better performance. Finally, task accuracy is measured by specific performance parameters that ensure the desired outcomes are met. \gls{ai} algorithms can be trained on \gls{eo} image datasets that often include images captured under varying conditions and featuring a small \gls{gsd} to achieve a high accuracy \cite{ijgi11080445}.


Several performance metrics are used to evaluate the goodness of an algorithm, with the supported \gls{fps} being crucial for real-time applications. 
Precision and recall refer to the proportion of correctly detected objects among all detected objects and among all actual objects, respectively. In addition, the \gls{map} (ranging between 0 and 1) is derived from the precision and recall values. Another metric is the Intersection over Union (IoU), which measures the overlap between the predicted boundary and the real object boundary (i.e., the ground truth). An IoU threshold is predefined (typical values being 0.5 and 0.95) to determine whether a detection is correct.



Data fusion provides a versatile data acquisition and processing solution to 
integrate the different data types, thereby improving the accuracy and reliability of the results \cite{Zhuang_multisensor_2023}. Furthermore, data fusion can employ data sourced from multiple satellites to provide a more comprehensive understanding of the observed phenomena while reducing power consumption and minimizing delays.
Multimodal data fusion can occur at different levels: observation-level data, corresponding to raw information captured by the sensors; feature-level data, which refers to vector embedding representations; and decision-level, which refers to the final verdict according to the task. As the fusion progresses through these levels, the complexity and semantic depth increase, leading to potential performance improvements. While classical signal processing has predominantly focused on unimodal observation-level fusion and decision-level fusion (see references in \cite{deniz2023semantic}), the field of multimodal fusion remains underexplored due to the inherent complexity of merging different data types. In this respect, \gls{ai} has emerged as a promising tool to address multimodal fusion at any level. Namely, each input is processed by a model (e.g., neural network), resulting in a vector embedding. Then, these representations are combined (e.g., appended) and further processed by another \gls{ai} model. Whether the embedding corresponds to the input itself, an intermediate representation or the final decision, the architecture can encompass multimodal data fusion at any level \cite{Zhao_DeepMultimodal_2024}. 

\section{LEO satellite constellations as an edge layer}



With the advent of softwarization of network functions, modern network edge elements oftentimes possess general-purpose processors. Edge computing 
exploits the processing resources at the network edge nodes to execute algorithms that operate either on user or local data. Edge computing reduces the latency when compared to cloud computing for user-initiated tasks, and enables traffic offloading and energy minimization using algorithmic compression~\cite{Ley23TCOM}. 

These benefits are particularly relevant in \gls{leo} satellites, often organized in constellations~\cite{Ley23TCOM}. The \gls{leo} satellite constellation can provide an edge layer to existing \gls{eo} satellites, or the satellites can be multi-purpose and do both the data acquisition and processing. Moreover, the density of the constellation determines the mode of operation and performance. Sparsely deployed \gls{leo} constellations have intermittent connectivity through the feeder links. In these cases, the latency with satellite edge computing can be as low as a few tens of milliseconds, whereas the latency with cloud computing can be in the order of a few hours, until the satellite can find a path towards the cloud server, requiring a high storage capacity at the satellite. If the constellation is densely deployed and \gls{isl}\cite{wang2022isl} are implemented, the satellites operate as a distributed edge computing architecture for distributed learning that avoids long propagation delays and increased traffic loads in the links towards the cloud servers. The feeder links and the \glspl{isl} can be based on conventional \gls{rf} technology, although there is a growing interest for \gls{fso} and hybrid \gls{fso}/\gls{rf} solutions.  

The conventional approach to \gls{eo} is sending the raw data to ground. This requires not only high capacity in the communication links but also storage at the space segment when the connectivity is intermittent. The feeder links, \gls{ul} and \gls{dl}, are particularly prone to congestion, since these are usually the links with the lowest capacity due to the movement of the satellites and the impact of atmospheric conditions. Satellite edge computing can alleviate congestion and expedite data processing in \gls{eo} by providing the computing capability to run the both classical and \gls{ai}-based algorithms on-board the satellites. Namely, images can be processed and compressed by the cooperating edge satellites, either with a classical algorithm, such as JPEG, or with more advanced semantic-empowered processing algorithm (e.g., object recognition and prediction), before being sent to the \gls{gs}.

\section{Semantic and goal-oriented \gls{eo}}

The term \gls{semcom}, as it is used today, encompasses the semantic and effectiveness levels of a communication system as originally defined by Shannon and Weaver~\cite{deniz2023semantic}. The effectiveness level is also known as goal- or task-oriented communications. Major efforts have been dedicated to precisely define what \gls{semcom} is, and to distinguish between the two problems, semantic and goal-oriented. However, the interplay between them makes it impractical to establish a generic boundary. Instead, it is more functional to define scenarios and optimization problems by their degree of semantic or goal-oriented relevance and, based on that, choose the proper set of requirements, performance metrics, and optimization methods. For example, from the three illustrative applications defined in the introduction, image reconstruction has a strong semantic component as the images are semantically compressed at the space segment and then interpreted at ground. The other two, real-time object detection and tracking, incorporate elements from both: a semantic feature extraction is embedded in a goal-oriented optimization, which aims at the correct and timely monitoring/tracking of the objects at the destination, rather than the individual image transmission.

\subsection{Architecture and general procedure}
The main elements of the architecture are described next. 

\noindent \textbf{\gls{gs}}: The ground segment is composed of one or few \gls{gs}, connected to end-users and to the space segment. It receives requests from the end-users, receives the results from the edge layer, and performs the semantic interpretation or inference. The \gls{gs} has also a strong (cloud) computing power that is used for non time-sensitive information and for complementing the edge capacity. Regarding communications, the feeder link connecting the \gls{gs} to the edge layer is often the capacity bottleneck.

\noindent \textbf{\gls{eo} satellites}: The \gls{eo} satellites perform the data acquisition, whose quality is impacted by the presence of atmospheric turbulence. The frame captured by a satellite camera covers a large surface area, e.g., $>170$ km$^2$ at altitude 617 km for WorldView-3. The frame is then split into smaller images characterized by the \gls{gsd}. If the \gls{eo} satellites are not part of the edge layer and have limited computing and storing capabilities, then must send the raw data (the images) to the edge layer to be processed and forwarded to ground, using \glspl{isl}.   

\noindent \textbf{Edge nodes}: The \gls{leo} satellite constellation provides edge computing capabilities for distributed processing and learning. Following \gls{semcom} principles, the semantic encoding/decoding leverages the \gls{kb}, which represents the shared knowledge with ground. Establishing the \glspl{kb} is a complex and time-consuming process,  done off-line at the \gls{gs},  that involves learning from the perceived environment, i.e., from large datasets. Then, the pre-trained algorithms are uploaded on-board.
With time, the \glspl{kb} diverges and might lead to wrong semantic inferences and interpretations, so \gls{kb} alignment and continual learning are required. The processed information is routed to the \gls{gs} through the constellation, which requires a routing algorithm. 


\noindent \textbf{Atmospheric, physical, and semantic noise}: The entire system is affected by various types of noise. Atmospheric noise or turbulence degrades satellite images, yet its impact is tolerable to some degree. Physical noise corrupting the transmissions over the \gls{isl} and the feeder links might cause bit errors, but can be largely corrected by channel decoding. Finally, semantic noise appears in the message interpretation processes due to, e.g., a mismatch between source and destination \gls{kb}.

\begin{figure}[t]
\centering
\includegraphics[width=3.5 in] {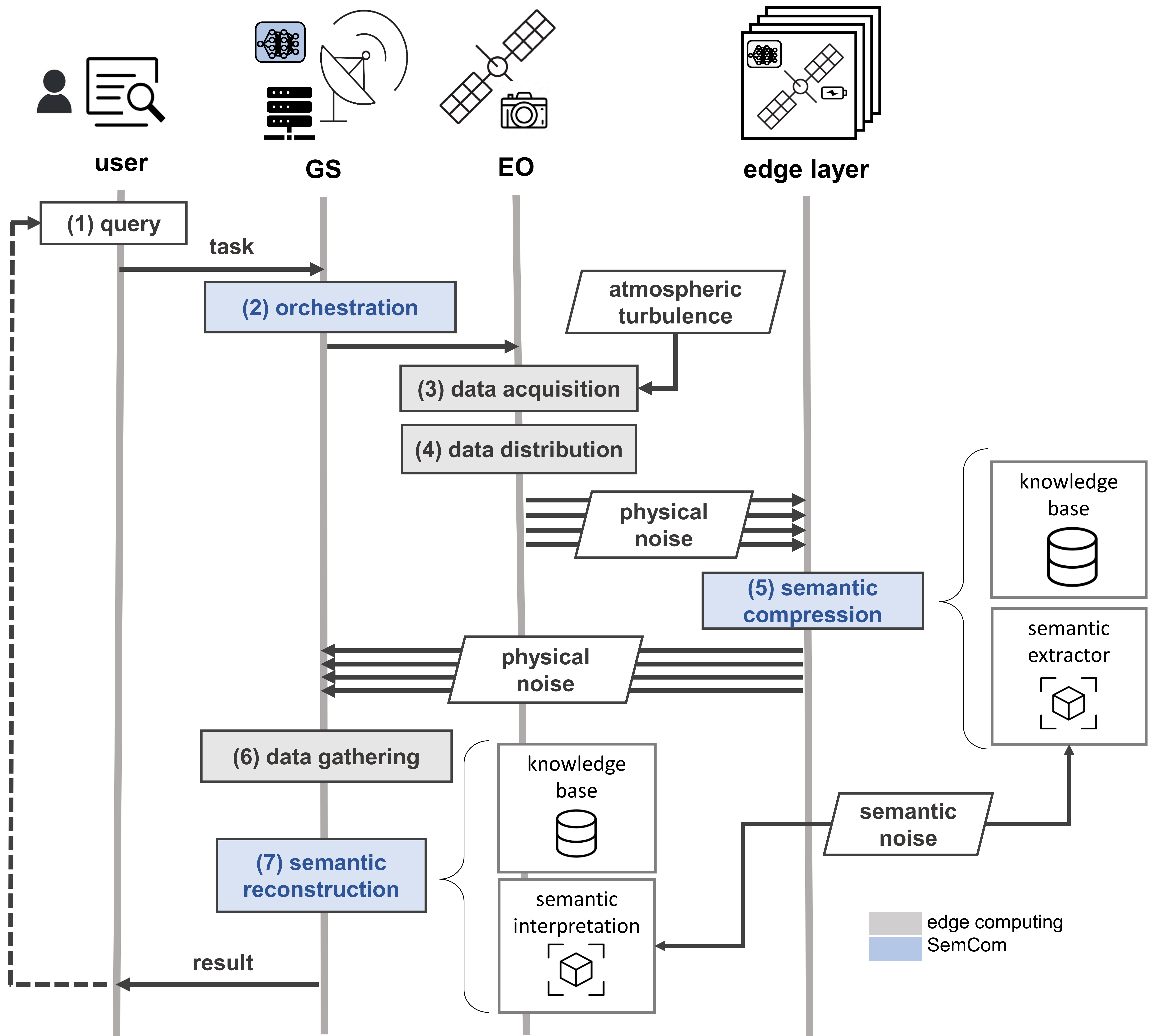}
\caption{Procedure of the proposed semantic and edge computing empowered \gls{eo}. } \vspace{-0.4cm} \label{fig:semanticarchitecture}
\end{figure}
This framework supports the general procedure illustrated in Figure~\ref{fig:semanticarchitecture}. An end-user query initiates a pull-based process in which the destination node requests information from the source. Specifically, the request is for the source node to sense the environment (i.e., take pictures) and transmit such updated information to the destination. The query can be semi-static (``Map area X for Y months'') or dynamic (``Track vessel with ID X'' or ``Map the area around volcano Y in eruption''). 
The query must be translated to a technical request to the network, in the form of an orchestration of the energy, computation, and communications resources, typically done at the \gls{gs}. This is a highly complex scheduling problem~\cite{eobook} which aims to maximize the observation profit for solving a specific task and using the minimum amount of resources. Depending on the nature of the query, the orchestration too will be more static or dynamic. Moreover, the orchestration must consider the current network conditions, e.g., the weather conditions, the possible subsystem failures, and the availability of resources. The \gls{eo} satellites acquire and distribute the data to the edge layer for the semantic parallel processing. The result from each cooperating edge satellite is sent to the \gls{gs} for data gathering and semantic reconstruction. If not all data is processed on the edge layer, due to timing constraints, then the rest of raw data is to be processed at the cloud \gls{gs}.
Based on the final semantic interpretation sent from the \gls{gs} to the end-user, a new query can be generated, closing the communication loop. Unlike a classical communication system, the objective of the communication and networking protocols (e.g., modulation and coding, schedulers, and routing) is not only that of increasing the reliability, for which classical link- and network- level metrics are used (e.g., throughput, bit error rate, end-to-end delay), but to fulfil the task, which is parameterized in semantic metrics like the ones defined next.  

\subsection{Metrics}
A fundamental aspect of \gls{semcom} is to find the set of relevant and practical metrics that quantify the gains of the reformulated resource optimization problems. The literature about this is huge (see e.g., \cite{deniz2023semantic}), but can be categorized as follows. 

\noindent  \textbf{Task-achievement metrics}: These metrics aim to capture the semantic meaning and usefulness of the communicated information for the intended task. 
When dealing with images, distortion-perception metrics can be used to evaluate how well the communication system preserves the relevant features of the images for accurate classification. 
The specific distortion metric used depends on the type of data, often involving squared error and Hamming distance. However, these metrics do not adequately capture semantic content. Instead, metrics that focus on the perceptual quality of the data are more suitable. For instance, in image processing, \gls{mse} can be used as a distortion metric, while the \gls{ssim} serves as a perceptual quality metric. For tasks like vessel detection, the primary focus is maximizing classification metrics such as recall.

\noindent \textbf{Time metrics}: 
For most applications, information must be received within a designated timeframe, with a maximum allowable delay. The delay is usually defined as the time elapsed from image capture to its reception at the ground station. Time performance is crucial and often closely tied to the level of task achievement. Nonetheless, we classify time as a separate category because of its central role in optimization of the resource allocation. Another consideration is that storage and queues have limited capacity. Ensuring the stability of processing and communication queues is vital to keeping the system operating efficiently without congestion. If each satellite can process and transmit data at a pace that exceeds the rate at which data arrives, the system is considered stable. Otherwise, the process is not stable and images risk to be discarded or not acquired. In this second case, we need to devise some strategies to reduce the rate of images acquired by the satellites.

\noindent \textbf{Energy metrics}: The amount of  energy available at the satellite is a critical constraint, as satellites rely upon solar panels for electrical power generation and its storage in the batteries for use during the eclipse periods. In a \gls{semcom} optimization, the focus is on the power consumed by the processing and communication subsystems. The energy consumption of the on-board and off-board algorithms is highly impacted by the processing architecture, which can be either \gls{cpu} or \gls{gpu}-based. \glspl{gpu} have massive parallel processing power, while \glspl{cpu} have fewer, more complex processing units optimized for general-purpose computing. The energy consumption of the data transmission through the \glspl{isl} and the feeder link depends on the used technology, \gls{rf} or \gls{fso}.


\subsection{Energy-time-accuracy trade-offs in \gls{semcom}}
 Semantic compression is adaptable to specific contexts and tasks, allowing for effective semantic interpretation of images at the receiver. In particular, in our case the \gls{kb} shared between satellites and \gls{gs} is utilized to achieve  compression levels some order of magnitude higher than traditional fixed methods. Performance may be affected by atmospheric turbulence and communication noise, which should be factored into the resource allocation strategy to prioritize satellites and images with higher decoding success rates. For real-time object tracking, the ground station estimates vessel trajectories in real-time, and resource allocation considers the reconstruction status. The complexity of resource allocation increases with closed-loop communication and satellite control based on specific tasks. Consider the case where the end user sends a query to the \gls{gs}, requesting information such as geographical coordinates or vessel IDs. The \gls{gs} translates this query into a directive for specific satellites to capture images. If the \gls{gs} cannot determine which satellites cover the area of interest, a general query is sent to all available satellites to capture new images. The request is then routed to the target satellites, where closed-loop control determines orbit and attitude parameters for data acquisition. This decision is influenced by network conditions (e.g., weather, available resources, subsystem failures) and the system status (e.g., tracked objects).
 
Accordingly, there is a delicate balance between accuracy, delay, and energy consumption when fulfilling EO requests. These conflicting objectives  must all  be considered in the resource allocation optimization, which can be conceptually formulated as follows: Identify the best set of \gls{eo} and edge satellites for image capturing, semantic extraction, encoding, and routing to the ground, with the aim of minimizing energy consumption while meeting accuracy and timing constraints. Energy consumption is influenced mainly by the on-board execution of algorithms and data transmission. As a result, semantic extraction has two opposing impacts: it increases energy consumption due to processing but decreases energy usage by reducing the amount of data that needs to be transmitted, since processed data is significantly smaller than raw data.
Additionally, the space segment must take into account the hard constraint of executing and completing algorithms before the next frame is captured and processed. If this is not achieved, images may need to be discarded or not captured to avoid congestion in the system. If only a portion of the data can be processed, the remaining data will be transmitted in its raw form to the \gls{gs}, thereby increasing the load on the communication network and prolonging transmission time.

\begin{figure}[t]
\centering

    \centering
    \includegraphics{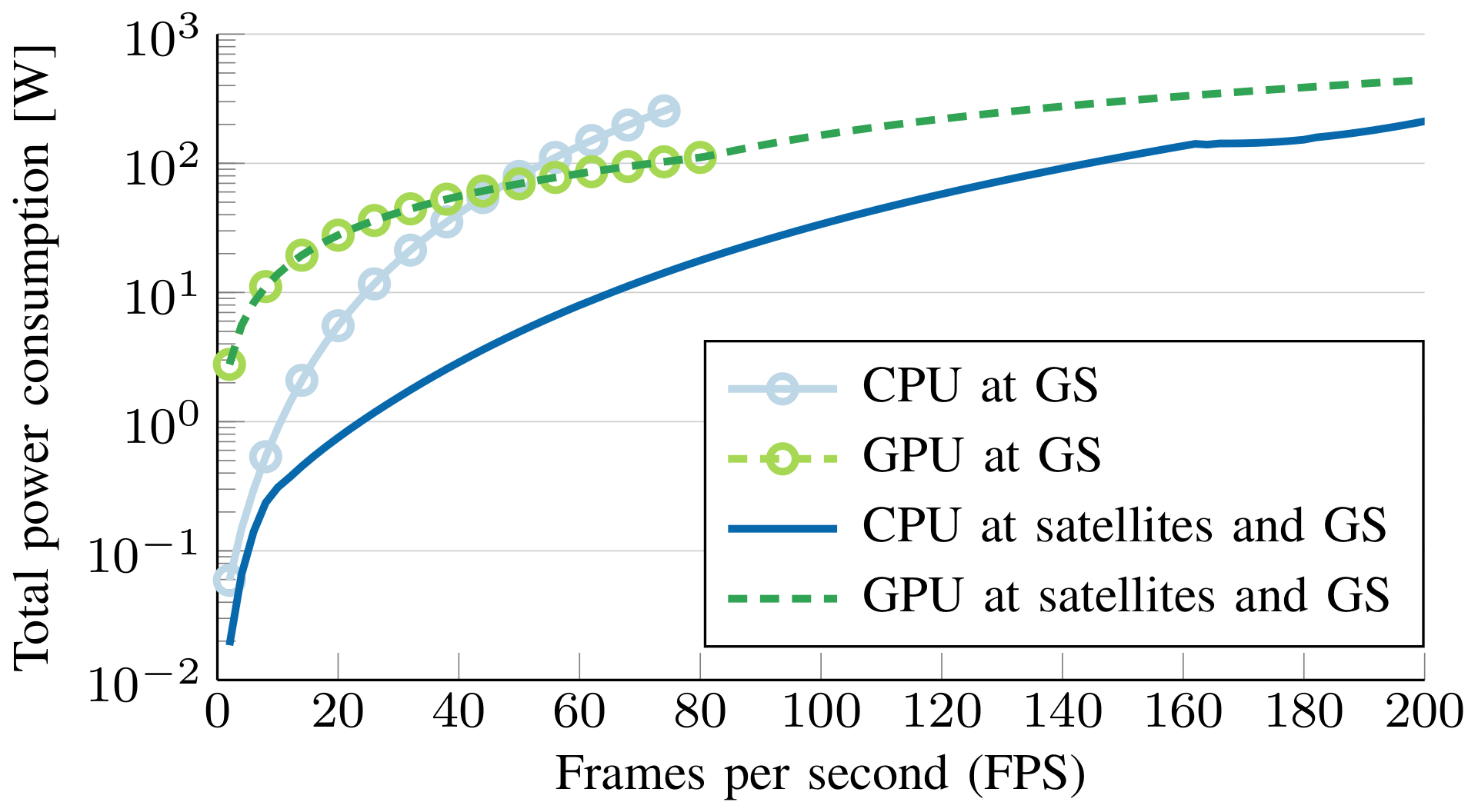}

\caption{Minimum power consumption in logarithmic scale versus number of \gls{fps} for \gls{cpu} and \gls{gpu} computing architectures. 
} \label{fig:powerconsumption_with_baselines}\vspace{-0.4cm} \label{fig:results}
\end{figure}

\section{Performance evaluation}

We provide illustrative results of the semantic optimization of the considered system, i.e., a satellite constellation with $23$ satellites that provides edge computing resources to an \gls{eo} satellite. The \glspl{isl} use \gls{fso} and the feeder link \gls{rf}~\cite{Ley23TCOM}. The task is to detect the ships located in an area of interest. Parameters from the satellite WorldView-3 \cite{worldview-13} are used for the image acquisition model. An \gls{eo} satellite captures a given number of \gls{fps} with a resolution of $600\times 600$\,pixels and a \gls{gsd} of $0.5$\,m, which might be affected by atmospheric turbulence. 
We consider YOLOv8 for ship detection, which requires $W_\text{YOLO}=79.1$ G floating-point operations (FLOPs) per image. We evaluate the performance with \gls{cpu} and \gls{gpu} architectures at the satellites and the \gls{gs} using statistical models of the execution time. Thus, each edge satellite has either an 8-core CPU operating at a maximum frequency of $1.8$\,GHz, or an NVIDIA T1000 GPU where YOLOv8 runs at $18.11$\,\gls{fps} 
with a power consumption of $50$\,W. The GS has either a 64-core 
CPU operating at a maximum frequency of $2.6$\,GHz or an NVIDIA Quadro RTX 5000 GPU where YOLOv8 runs at $80.812$\,\gls{fps}  
with a power consumption of $110$\,W.  
On average, the size of the semantic data (i.e., ships and bounding boxes) is $336$ bits per image.



Figure \ref{fig:powerconsumption_with_baselines} shows the minimum power consumption versus the \gls{fps} as the result of the energy-time-accuracy optimization problem. The two baselines are conventional cloud computing at the \gls{gs} with the two architecture options, \gls{cpu} and \gls{gpu}. With a low \gls{fps}, processing the images at the \gls{cpu} results in lower power consumption. The main reason for this is that the \glspl{cpu} can be configured to operate at the optimal \gls{cpu} frequency~\cite{Ley23TCOM}, while the \glspl{gpu} operate at maximum frequency. As the \gls{fps} grows, processing the images at the \gls{gs} \gls{cpu} becomes the worst alternative and using a combination of edge and cloud processing becomes necessary to achieve the minimal power consumption while meeting the optimization constraints. 



\section{Conclusions}
While \gls{semcom} has been a hot topic in recent years, its application in space is yet to be explored. Considering the huge volume of data generated by \gls{eo} missions, we have investigated the potential of using \gls{semcom} together with the edge computing power provided by a \gls{leo} satellite constellation to support real-time object localization and detection. This combination 
expedites  object recognition and prediction, and minimizes data transfer over congested links without jeopardizing the performance. The proposed framework supports a great diversity of semantic and goal-oriented problems, illustrated in exemplary results with real images and satellite mission parameters.

\bibliography{refs}
\bibliographystyle{IEEEtran}




\end{document}